\documentclass[aps,prb,amsmath,amssymb,rsi,groupedaddress,twocolumn]{revtex4}
\usepackage{graphicx}
\usepackage{epsfig}
\usepackage{amsmath}
\usepackage{graphics}
\usepackage{epsfig}
\usepackage{amsmath}
\usepackage{amsfonts}
\usepackage{amssymb}
\usepackage{xcolor}
\allowdisplaybreaks

\begin{document}

\title{A model for  dynamical solvent control of  molecular  junction electronic properties} 
\author{Maxim F. Gelin\footnote{E-mail: maxim@hdu.edu.cn}} 
\affiliation{School of Sciences, Hangzhou Dianzi University, 310018 Hangzhou, China}
\author{Daniel S. Kosov\footnote{E-mail: daniel.kosov@jcu.edu.au}} 
\affiliation{College of Science and Engineering, James Cook University, Townsville, QLD, 4811, Australia}

\begin{abstract}
Experimental measurements of electron transport properties of molecular junctions are often performed in solvents. Solvent-molecule coupling and physical properties of the solvent can be used as the external stimulus to control electric current through a molecule. 
In this paper, we propose a model, which includes dynamical effects of solvent-molecule interaction in the non-equilibrium Green's function calculations of electric current.
The solvent is considered as  a macroscopic dipole moment that reorients stochastically and interacts with the electrons tunnelling  through the molecular junction. The Keldysh-Kadanoff-Baym equations  for electronic Green's functions are solved in time-domain with subsequent averaging over  random realisations of rotational variables using Furutsu-Novikov method  for exact closure of infinite hierarchy of stochastic 
correlation functions. {{}  The developed theory requires the use of wide-band approximation  as well as classical treatment of solvent degrees of freedom.}
The theory is  applied to a model molecular junction.  It is demonstrated that not only electrostatic interaction  between molecular junction and solvent but also solvent viscosity can be used to control electrical properties of the junction. Aligning of the rotating dipole moment breaks particle-hole symmetry of the transmission favouring either hole or electron transport channels depending  upon the aligning potential.
\end{abstract}

\maketitle

\section{Introduction}
The single-molecule electrical experiments now go way beyond the initial current as a function of applied voltage measurements.
The  environment is no longer playing a passive role but used as experimental means to control electronic properties of molecular electronic  junction.\cite{doi:10.1021/acs.chemrev.5b00680}
Mechanical stretching, $\pi$ stacking,\cite{Wu:2008aa} hydrogen bonding\cite{doi:10.1021/ja311463b} and supramolecular interactions \cite{doi:10.1021/ja312019p} have been used to change electrical properties of molecular junctions.
Recently, there is a growing interest in the use of  solvent to control  transport of electrons in single-molecule junction. \cite{kuznetsov-ndr,doi:10.1021/acs.jpcc.5b08877,solvent12,PhysRevLett.102.086801,PhysRevLett.78.4410,solvent2016,Kotiuga:2015aa,doi:10.1021/nl200324e,doi:10.1021/acs.jpcc.5b06867,kuznetsov-ndr,C5SC02595H,Kornyshev6799}

The scope of theoretical studies of  solvated molecular electronic junctions is somewhat  limited. On the computational side,  density functional theory nonequilibrium Green's function based  calculations  take into account   a few surrounding solvent molecules  kept in the fixed optimised geometry to mimic the effect of molecule-solvent interaction.\cite{PhysRevLett.102.086801,doi:10.1021/acs.jpcc.5b08877,doi:10.1021/acs.jpcc.5b06867}
Several simple models were developed to treat electrolyte environment of molecular electronic junctions via Poisson-Boltzmann equations incorporating effect of the solvent into  voltage drop across the  junction  with subsequent rate equation calculations of electric current.\cite{Kornyshev6799,doi:10.1021/cr068073+}
Born solvation model was combined  in the self-consistent manner with static nonequilibrium Green's functions calculations electric current calculations to treat the problem.\cite{solvent2016}

These theoretical works provide an important insight on the role of the  solvent,  however, they all assumed that the time-scales for electrons and solvent motion can be completely separated. In other words, all time-dependent effects from solvent response and rearrangement due to the dynamical charging and discharging of the molecular junction by current-carrying electrons have been neglected.  In this paper, we developed the model which takes into account solvent dynamics.   The solvent is modelled as  a macroscopic, stochastically rotating  dipole moment which interacts with the electrons tunnelling  through the molecular junction.  This interaction makes the Hamiltonian for quantum transport problem explicitly time-dependent
which is handled exactly by nonequilibrium Green's functions calculations. The evaluation of observables  obtained from electronic Green's functions requires  the averaging over realisations of Gaussian stochastic process, which is performed using Furutsu-Novikov method for stochastic calculus.\cite{klyatskin2005}
As far as time-evolution of the system is concerned, quantum electronic dynamics are treated on an equal footing with classical solvent dynamics in our method.

The paper is organised as follows.  Section II describes the theory. It introduces  model Hamiltonian for molecular junctions, main definitions for Green's functions and self-energies, as well as describes solution of the Keldysh-Kadanoff-Baym equations coupled to the stochastic rotor dynamics. Section II also provides expression for electronic current averaged over stochastic realisations of solvent dynamics and details  the use of Furutsu-Novikov method.
Section III illustrates the theory by calculations on model molecular systems and discusses the main physical observations. 
The main results of the paper are summarised in section IV.  The technical details of the derivations and numerical algorithm description are relegated to the appendices.

\section{Theory}

\subsection{Model Hamiltonian }

In this section we discuss the physical model for molecular electronic junction coupled to solvent.   The system  Hamiltonian consists of five parts
\begin{multline}
H(t) = H_{\text{molecule}} + H_{\text{leads}} + H_{\text{solvent}}(t) \\
+ H_{\text{leads-molecule}} + H_{\text{solvent-molecule}}(t).
\end{multline}
Here $H_{\text{molecule}}$ is the molecular Hamiltonian,  $H_{\text{leads}}$ describes electron reservoirs in the right and left leads. Coupling between the molecule  and the leads  is given by $H_{\text{leads-molecule}}$. The solvent  is described by  $H_{\text{solvent}}$ and the coupling between molecule and the solvent is denoted by $H_{\text{solvent-molecule}}$, these parts of the Hamiltonian are time-dependent.

The molecule is modelled as a single resonant-level, which can be occupied by the zero or one spin-less electrons (the electron spin will not be considered explicitly in our derivations but we will recreate the factor of two due to spin degeneracy in the final expressions for electric current). The corresponding Hamiltonian is 
\begin{equation}
H_{\text{molecule}}=\epsilon d^\dag d,
\end{equation}
where  $d^\dag$ ($d$) creates (annihilates) electron on molecular orbital with  energy $\epsilon$.

The left and right leads  are modelled as macroscopic reservoirs of non-interacting electrons
\begin{equation}
H_{\text{leads}}  =  \sum_{k \alpha} \epsilon_{k \alpha} d^\dag_{k \alpha} d_{k \alpha},
\end{equation}
where $d^\dagger_{k \alpha}$ and $d_{k \alpha}$ are the creation and annihilation operators for a single particle state of energies $\epsilon_{k \alpha}$ for  left $\alpha=L$ or right $\alpha=R$ leads. The  couplings between leads and molecule are described by the tunnelling interaction
\begin{equation}
H_{\text{leads-molecule}}  =  \sum_{k \alpha} ( t_{k \alpha}   d^\dag_{k \alpha} d  +  t^*_{k \alpha}  d^\dag d_{k \alpha}), 
\end{equation}
where $ t_{k \alpha}$ is the tunnelling amplitude between molecule and leads states.

The  solvent, which is  modelled as classical rigid rotating dipole moment  $\mathbf \mu$ with moment of inertia $I$, is described by the time-dependent Hamiltonian
\begin{equation} \label{HS}
H_{\text{solvent}}(t)= \frac{1}{2 I} \left( \frac{d \phi}{d t} \right)^2,
\end{equation}

where $\phi$ is the orientation angle of the dipole moment relative to the molecule.  {{}  This level of description of  solute-solvent interactions is common in theory of rotational relaxation in polar solvents and constitutes the essence of stochastic cage model \cite{coffey}, where the solvent shell surrounding the molecule is described as an effective quasiparticle which interacts with the molecule by appropriate anisotropic potential and stochastically interacts with the balk solvent.   We thus assume }  that the orientation angle  $\phi$  and  the conjugated
rotational frequency $\Omega $ obey the stochastic evolution equations 
\begin{equation}
\frac{d\phi(t)}{dt}=\Omega (t),\label{s1}
\end{equation}
and
\begin{equation}
\frac{d\Omega (t)}{dt}=G(\phi(t))-\xi\Omega (t)+F(t).\label{s2}
\end{equation}
Here $U(\phi)$
is an external potential, 
\begin{equation}
G(\phi)=-\frac{dU(\phi)}{d\phi}
\end{equation}
is the torque acting on the rotator, and $F(t)$ is a $\delta$-correlated
stochastic Gaussian process, 
\begin{equation}
\langle F(t)\rangle=0,\,\,\,\langle F(t)F(t')\rangle=2\xi\delta(t-t'),
\label{FF}
\end{equation}
where $\xi$ is the rotational friction. 
{{}  The fluctuation-dissipation theorem relation between the noise and viscosity is assumed here, which implies that the solvent 
is maintained in thermodynamic equilibrium.}
We use the dimensionless variables,
in which $\Omega $ and $\xi$  are expressed in units
of $\sqrt{k_{B}T/I}$, where 
 $k_{B}$ is Boltzmann's constant and  $T$ is a temperature.

The interaction between the molecule and the solvent is given by
point charge-dipole interaction
\begin{equation}\label{HSM}
H_{\text{solvent-molecule}}(t)= \frac{1}{4 \pi \epsilon_0}  \frac{ e d^\dag d   \; \mu \cos (\phi)}{r^2},
\end{equation}
where the instantaneous charge of the molecule due to tunnelling of electrons is given by electron number operator
$d^\dag d$ multiplied by electron charge $e$, $\mu$ is magnitude of the classical dipole moment and $r$ is the distance between molecule and   {{}  solvent cage centre of mass, and  $\epsilon_0$ is electrical permittivity}. As typical in electron transport calculation we assume that $e$ carries the negative sign ($e= -|e|$).
This part of the Hamiltonian is explicitly time-dependent due to angle $\phi$ being time-dependent variable undergoing stochastic fluctuations.
{{} 
	Of course, such a treatment of solvent-molecule interaction is an oversimplification, but it allows us to conveniently rewrite Eq. (\ref{HSM}) as   
	\begin{equation}\label{HSMa}
	H_{\text{solvent-molecule}}(t)= - \lambda \cos (\phi),
	\end{equation}
where the parameter 
\begin{equation}
\lambda = -\frac{1}{4 \pi \epsilon_0} \frac{e \mu}{r^2}.
\end{equation}
can be interpreted as an effective solvent-molecule coupling strength.  Eqs. (\ref{HS})-(\ref{HSMa}) describe a variant of the stochastic cage model which is called the itinerant oscillator model \cite{coffey}. Despite its simplicity, this model captures essential physics. It  
was successfully applied to the description of molecular rotation in polar liquids and solvents.

It is convenient to include solvent-molecule interaction into the molecule Hamiltonian by introducing time-dependent energy level
\begin{equation}
h (\phi(t))= \epsilon - \lambda \text{cos} (\phi(t)). 
\label{epsilont}
\end{equation}
Then combining two terms  together we  get
\begin{equation}
H_{\text{molecule}} + H_{\text{solvent-molecule}}= h (\phi(t)) d^\dag d.
\end{equation}
}

\subsection{Green's functions and self-energies}
In this section, Green's functions and self-energies definitions are introduced and basic notation is established; they will be used throughout the paper.
The exact and non-adiabatic (computed along a given angular trajectory $\phi (t)$) retarded, advanced and lesser Green's functions in the molecular space are defined as \cite{haug-jauho}
\begin{equation}
\mathcal{G}^R(t,t') = -i \theta(t-t') \langle \{d(t), d^\dag(t')\} \rangle,
\end{equation}
\begin{equation}
\mathcal{G}^A(t,t') = \Big({\cal G}^R(t',t) \Big)^*
\end{equation}
and
\begin{equation}
\mathcal{G}^<(t,t') = i \langle d^\dag (t') d (t) \rangle.
\end{equation}
The retarded, advanced and lesser self-energies for $\alpha$ lead are given by the standard expressions
\begin{equation}
{\Sigma}_{\alpha}^R(t,t') = -i {{}  \theta(t-t')} \sum_{k} t^*_{k \alpha}  e^{-i \epsilon_{k} (t-t')} t_{k \alpha},
\end{equation}

\begin{equation}
{\Sigma}_{\alpha}^A(t,t') = \Big( {\Sigma}_{\alpha}^R(t',t) \Big)^*,
\label{sigmaAt}
\end{equation}
\begin{equation}
{\Sigma}_{\alpha}^<(t,t') =2 \pi i \sum_{k} t^*_{k \alpha}  f_{\alpha}(\epsilon_k) e^{-i \epsilon_{k} (t-t')} t_{k \alpha}.
\end{equation} 
Here  $f_{\alpha}$ is the Fermi-Dirac distribution. The total self-energies are the sum of contributions from the left and right leads as
\begin{equation}
{\Sigma}^{R,A,<}(t,t') =  {\Sigma}^{R,A,<}_{L}(t,t')+ {\Sigma}_{R}^{R,A,<}(t,t').
\end{equation}

The self-energies in energy-domain are defined as a Fourier transformation
\begin{equation}
{\Sigma}^{R,A,<}(\omega ) = \int d(t-t') e^{i \omega (t-t')} {\Sigma}^{R,A,<}(t,t').
\end{equation}
The wide-band approximation will be used in our derivations and calculations: within this approximation the time-dependence of retarded and advanced self-energies is reduced to delta-function
\begin{equation}
\Sigma_{\alpha}^R(t,t')=  -\frac{i}{2} \Gamma_{\alpha} \delta(t-t'), \;\;\; \Sigma_{\alpha}^A(t,t' )=  \frac{i}{2} \Gamma_{\alpha}\delta(t-t'), 
\end{equation}
where the level-broadening function $\Gamma_\alpha$ is time-independent parameter which describes the strength of molecule-lead coupling.  {{}  The wide-band approximation means the assumption that both the leads' density of states and  all tunneling amplitudes between leads' and molecular states are energy-independent constants. Our approach depends critically on the use of the wind-band approximation to solve analytically Keldysh-Kadanoff-Baym equations.}

Notice that we set $\hbar=1$ in all definitions of Green's functions. This implies that all quantum mechanical energy related quantities $\epsilon$, $\lambda$, and $\Gamma_{\alpha}$ are scaled by $\hbar$ and measured in the units of frequency (1/time).

\subsection{Solution of Keldysh-Kadanoff-Baym equations and expression for the current}
Suppose that a stochastic trajectory of the dipole rotation angle $\phi(t)$ is known from the solution of the Langevin equation. Our goal in this section is to solve exactly Keldysh-Kadanoff-Baym equations for the Green's functions and consequently obtain a concise expression for electric current averaged over the stochastic trajectory $\phi(t)$.

We begin with the general expression for electric current from $\alpha$ lead  in time-dependent system\cite{haug-jauho}
\begin{multline}
J_\alpha(t) = 2 e \text{ Re } \int_{-\infty}^{+\infty} dt' \Big[ \mathcal{G}^<(t,t') \Sigma_\alpha^A(t',t) \\
+ \mathcal{G}^R(t,t') \Sigma_\alpha^<(t',t) \big].
\end{multline}
Next, we transform self-energies to the energy domain whilst leaving molecular Green's functions time-dependent
\begin{multline}
J_\alpha(t) = 2 e \text{ Re }  \int_{-\infty}^{+\infty} \frac{d \omega }{2 \pi}   \int_{-\infty}^{+\infty} dt'  e^{-i \omega (t'-t)}  
\\
\times \Big[ \mathcal{G}^<(t,t') \Sigma_\alpha^A(\omega ) + \mathcal{G}^R(t,t') \Sigma_\alpha^<(\omega ) \big].
\end{multline}

Electric current should satisfy the continuity equation at each time moment $t$
\begin{equation}
\frac{d N}{dt} = J_L(t) + J_R(t),
\end{equation}
where $N$ is the total number of electrons in the molecule at time $t$. Using the continuity equation and  rearranging the terms, we  eliminate lesser Green's function from the expression and get
\begin{widetext}
\begin{equation}
J_L(t) = \frac{\Gamma_L}{\Gamma}  \Big\{\frac{d N}{dt} 
 -  2 e  \Gamma_R   
  \text{ Im }  \int_{-\infty}^{+\infty} \frac{d \omega }{2 \pi}   \int_{-\infty}^{+\infty} dt' 
 e^{-i \omega (t'-t)} \mathcal{G}^R(t,t') [f_L(\omega ) - f_R(\omega )] \Big\}.
 \end{equation}
\end{widetext}
Let us  average the above equation over the stochastic realisation of the dipole moment rotation process. The average 
\begin{equation}
\langle  \frac{d N}{dt}  \rangle =0 
\end{equation}obviously disappears, otherwise the molecule would either  accumulate or loose charge continuously.  The averaged current becomes
\begin{widetext}
\begin{equation}
\langle J_L \rangle =
 -  2 e \frac{\Gamma_L \Gamma_R}{\Gamma}  \\  \text{ Im }  \int_{-\infty}^{+\infty} \frac{d \omega }{2 \pi}   \int_{-\infty}^{+\infty} dt'  e^{-i \omega (t'-t)}  
 \langle \mathcal{G}^R(t,t') \rangle  [f_L(\omega ) - f_R(\omega )].
 \label{JL-av}
\end{equation}
\end{widetext}

Notice, that the averaged electric current does not depend on time, since the retarded Green's function will depend on relative time only once averaged over the stochastic realisations.

Our next goal is to find the explicit expression for the retarded Green's function to enter it into the expression to electric current. We begin with the {{}  Keldysh-}Kadanoff-Baym equation 
for the retarded Green's function\cite{haug-jauho}
\begin{multline}
(i \partial_t - h(t))\mathcal{G}^R(t,t') \\
- \int_{-\infty}^{+\infty} dt_1 \Sigma^R(t,t_1) \mathcal{G}^R(t_1,t')
= \delta(t-t').
\end{multline}
It is reduced in the wide-band approximation to
\begin{multline}
(i \partial_t - h(t))\mathcal{G}^R(t,t') 
-\frac{i}{2} \Gamma \mathcal{G}^R(t_1,t') 
 = \delta(t-t'),
\end{multline}
where $h(t)$  denotes $h(\phi(t))$  given by equation (\ref{epsilont}) and 
\begin{equation}
\Gamma= \Gamma_L + \Gamma_R
\end{equation}
is the total level-broadening function. 
This differential equation can be solved analytically and yields
\begin{equation}
\mathcal{G}^R(t,t') = -i {{}  \theta(t-t')} e^{-\frac{1}{2} \Gamma (t-t')} e^{- i \int^t_{t'} dt_1 h(t_1)}
\label{gr}
\end{equation}
{{}  Notice that  analytical expression (\ref{gr}) requires the wide-band approximation treatment for leads' self-energies.}
Substituting retarded Green's function (\ref{gr}) into the expression for current (\ref{JL-av}) gives
\begin{widetext}
\begin{equation}
\langle J_L \rangle =
   2 e \frac{\Gamma_L \Gamma_R}{\Gamma}   \text{ Re }  \int_{-\infty}^{+\infty} \frac{d \omega }{2 \pi}   \int_{-\infty}^{t} dt'  e^{-i (\omega -\frac{i}{2} \Gamma) (t-t')}  \langle e^{- i \int^t_{t'} dt_1 h(t_1)}  \rangle [f_L(\omega ) - f_R(\omega )].
 \label{JL-av1}
\end{equation}
Changing variables of integration to $\tau=t-t'$ and taking into account that quantity $\langle e^{- i \int^t_{t'} dt_1 h(t_1)}  \rangle$ depends on relative time only once averaged over realisations of the stochastic process we arrive to 
\begin{equation}
\langle J_L \rangle =
   2 e \frac{\Gamma_L \Gamma_R}{\Gamma}   \text{ Re }  \int_{-\infty}^{+\infty} \frac{d \omega }{2 \pi}   \int^{\infty}_{0} d\tau  e^{-i (\omega -\frac{i}{2} \Gamma) \tau)} R(\tau) [f_L(\omega ) - f_R(\omega )]
 \label{JL-av2}
\end{equation}
where 
\begin{equation}
R(\tau) = \langle e^{- i \int^\tau_{0} dt_1 h(t_1)}  \rangle 
\label{Rt}
\end{equation}

\end{widetext}

Eq. (\ref{JL-av2}) can be rewritten in the Landauer form
\begin{equation}
\langle J_L \rangle = e
 \int_{-\infty}^{+\infty} \frac{d \omega  }{2 \pi}   T(\omega)  [f_L(\omega ) - f_R(\omega )]
 \label{JL-av3}
\end{equation}
 where we have introduced "transmission coefficient"
\begin{equation}
T(\omega) =    2 \frac{\Gamma_L \Gamma_R}{\Gamma}   \text{ Re }  \int^{\infty}_{0} d\tau  e^{-i (\omega -\frac{i}{2} \Gamma) \tau)} 
R(\tau).
\label{trans}
\end{equation}
 Notice that although being called "transmission coefficient", $T(\omega)$ given by  (\ref{trans}) should not be assigned  the meaning of  probability for electron with energy $\omega$ to tunnel across the molecule,  since the inelastic processes due to the dynamical coupling of tunnelling electron with rigid rotator are included into our consideration.

 \subsection{Averaging over the stochastic solvent dynamics}

The  aim here is to develop a method to compute $R(t)$ of Eq. (\ref{Rt}) 
(for convenience of notation, $t$ will be used as a variable in $R(t)$) 
where  averaging $\langle ... \rangle$  is performed over realisations of stochastic variable $\phi$ and time-evolution of $\phi(t)$ is given by stochastic differential equations (\ref{s1}) and (\ref{s2}). 

Let us define the quantity    
\begin{equation} \label{r0}
 A_{t}([\phi(\tau)]) =e^{-i\int_{0}^{t}dt_1h(\phi(t_1))}.  
 \end{equation}
Hereafter, the notation $A_{t}[\phi(\tau)]$ means that $A$ is
a function of $t$ and a functional of the stochastic process $\phi(\tau)$. Evidently, 
\begin{equation} 
R(t)= \langle A_{t}([\phi(\tau)]) \rangle.
\end{equation}
Let us now introduce the stochastic functional 
\begin{equation}
R(\phi,\Omega ,t)=\delta(\phi-\phi(t))\delta(\Omega -\Omega (t))A_{t}[\phi(\tau)]\label{rr1}
\end{equation}
where $\phi(t)$ and $\Omega (t)$ are certain realisations of the
stochastic processes (\ref{s1}) and (\ref{s2}). Averaging Eq. (\ref{rr1}) over 
all realisations of these processes yields  
\begin{equation}
\rho(\phi,\Omega ,t)=\left\langle R(\phi,\Omega ,t)\right\rangle, \label{r1}
\end{equation}
which can be interpreted as a probability density of $ A_{t}[\phi(\tau)] $
to have a certain value given $\phi(t)$ and $\Omega (t)$ are equaled
to $\phi$ and $\Omega $ at a time moment $t$. Then, according to van Kampen lemma,\cite{vanKampen} we obtain   
\begin{equation}
R(t) =\int_{-\infty}^{\infty}d\Omega \int_{0}^{2\pi}d\phi\rho(\phi,\Omega ,t).\label{r2}
\end{equation}

We are in the position now to derive a closed-form Fokker-Planck equation for  $\rho(\phi,\Omega ,t)$.
Formally differentiating $R(\phi,\Omega ,t)$ with respect to time,
using stochastic equations (\ref{s1}) and (\ref{s2}) and taking
the average, we obtain 
\begin{multline}
\partial_{t}\rho(\phi,\Omega ,t)=\Big(-ih(\phi)-\Omega \partial_{\phi}-G(\phi)\partial_{\Omega } 
+\xi\partial_{\Omega }\Omega \Big)\rho(\phi,\Omega ,t) \\
-\partial_{\Omega }\Omega \left\langle F(t)R(\phi,\Omega ,t)\right\rangle .\label{ra}
\end{multline}
Employing the Furutsu-Novikov formula\cite{klyatskin2005}, we get 
\begin{equation}
\left\langle F(t)R(\phi,\Omega ,t)\right\rangle =\intop_{0}^{t}dt'\left\langle F(t)F(t')\right\rangle \left\langle \frac{\delta R(\phi,\Omega ,t)}{\delta F(t')}\right\rangle .\label{FR}
\end{equation}
To evaluate the functional derivative, we follow the method of
Ref. \cite{klyatskin2005}:
\begin{multline}
\frac{\delta R(\phi,\Omega ,t)}{\delta F(t')}=\Big(-\partial_{\phi}\frac{\delta\phi(t)}{\delta F(t')}-\partial_{\Omega }\frac{\delta\Omega (t)}{\delta F(t')}
\\
-i\int_{0}^{t}dt''\frac{\delta h(\phi(t''))}{\delta F(t')}\Big)\delta(\phi-\phi(t))\delta(\Omega -\Omega (t))A_{t}[\phi(\tau)]
\\
=\Big(-\partial_{\phi}\frac{\delta\phi(t)}{\delta F(t')}-\partial_{\Omega }\frac{\delta\Omega (t)}{\delta F(t')}
\\
+i\lambda\int_{0}^{t}dt''\sin(\phi(t''))\frac{\delta\phi(t'')}{\delta F(t')}\Big)\delta(\phi-\phi(t))\delta(\Omega -\Omega (t))A_{t}[\phi(\tau)].
\label{FR1}
\end{multline}
The stochastic equations (\ref{s1}) and (\ref{s2}) can be rewritten in the integral form as follows: 
\begin{equation}
\phi(t)=\phi(0)+\intop_{0}^{t}d\tau\int_{-\infty}^{\infty}d\Omega \Omega \delta(\Omega -\Omega (\tau)),\label{s1a}
\end{equation}
and
\begin{widetext}
\begin{multline}
\Omega (t)=\Omega (0)+\intop_{0}^{t}d\tau\left(G(\phi(\tau))+\xi\Omega (\tau)+F(\tau)\right)=
\\
\Omega (0)+\intop_{0}^{t}d\tau\int_{-\infty}^{\infty}d\Omega \int_{0}^{2\pi}d\phi\left(G(\phi)+\xi\Omega +F(\tau)\right)\delta(\phi-\phi(\tau))\delta(\Omega -\Omega (\tau)).\label{s2a}
\end{multline}
Differentiating (\ref{s1a}) and (\ref{s2a}) with respect to $F(t')$,
we obtain:
\begin{equation}
\frac{\delta\phi(t)}{\delta F(t')}=-\intop_{t'}^{t}d\tau\int_{-\infty}^{\infty}d\Omega \Omega \partial_{\Omega }\delta(\Omega -\Omega (\tau))\frac{\delta\Omega (\tau)}{\delta F(t')},\label{s1b}
\end{equation}
\begin{multline}
\frac{\delta\Omega(t)}{\delta F(t')}=
\intop_{0}^{t}d\tau\int_{-\infty}^{\infty}d\Omega \int_{0}^{2\pi}d\phi\left[\frac{\delta F(t)}{\delta F(t')}-\left(G(\phi)+\xi\Omega +F(\tau)\right)\left(\partial_{\phi}\frac{\delta\phi(t)}{\delta F(t')}+\partial_{\Omega }\frac{\delta\phi(t)}{\delta F(t')}\right)\right]\delta(\phi-\phi(\tau))\delta(\Omega -\Omega (\tau))
\\
=1-\intop_{t'}^{t}d\tau\int_{-\infty}^{\infty}d\Omega \int_{0}^{2\pi}d\phi\left[\left(G(\phi)+\xi\Omega +F(\tau)\right)\left(\partial_{\phi}\frac{\delta\phi(t)}{\delta F(t')}+\partial_{\Omega }\frac{\delta\phi(t)}{\delta F(t')}\right)\right]\delta(\phi-\phi(\tau))\delta(\Omega -\Omega (\tau)).
\label{s2b}
\end{multline}
\end{widetext}
Due to causality, we have replaced 
\begin{equation}
\intop_{0}^{t}d\tau\rightarrow\intop_{t'}^{t}d\tau\label{Rep}
\end{equation}
because 
\begin{equation}
\frac{\delta\phi(t)}{\delta F(t')}=\frac{\delta\Omega (t)}{\delta F(t')}=0\,\,\,\mathrm{for}\,\,\tau<t'.\label{caus}
\end{equation}
Since the stochastic torque $F(t)$ is delta-correlated (Eq. (\ref{FF})),
we need to evaluate the functional derivative at $t'=t.$ Putting
$t'=t$ in Eqs. (\ref{s1b}) and (\ref{s2b}), we obtain 
\begin{equation}
\frac{\delta\phi(t)}{\delta F(t)}=0,\,\,\,\frac{\delta\Omega (t)}{\delta F(t)}=1.
\end{equation}
Formula (\ref{caus}) allows us to make the replacement
$\int_{0}^{t}dt''\rightarrow\int_{t'}^{t}dt''$ in Eq. (\ref{FR1}),
which insures that this integral vanishes at $t'=t$. Hence 
\begin{equation}
\frac{\delta R(\phi,\Omega ,t)}{\delta F(t)}=-\partial_{\Omega } \Omega  R(\phi,\Omega ,t).
\end{equation}
Finally, evaluating the integral in Eq. (\ref{FR}) through the explicit
expression (\ref{FF}) for $\left\langle F(t)F(t')\right\rangle ,$
we arrive at the Fokker-Planck equation 
\begin{multline}
\partial_{t}\rho(\phi,\Omega ,t)=\Big(-ih(\phi)-\Omega \partial_{\phi}-G(\phi)\partial_{\Omega }\\
+\xi\left(\partial_{\Omega }\Omega +\partial_{\Omega }^{2}\right)\Big)\rho(\phi,\Omega ,t).\label{FP}
\end{multline}
{{} 
 The first three terms in Eq. (\ref{FP}) describe evolution of $\rho(\phi,\Omega ,t)$ without action of the solvent, while the impact of the solvent is taken care of by the addition of the Fokker-Planck dissipation operator. A simple and intuitively clear  additive structure of Eq. (\ref{FP}) is caused by the Markovianity of the stochastic process $\{\phi(t),\Omega(t)\}$.}
Eq. (\ref{FP}) should be solved with the initial condition 
\begin{equation}
\rho(\phi,\Omega ,0)=\rho_{B}(\Omega )\rho_{B}(\phi)
\end{equation}
where 
\begin{equation}
\rho_{B}(\Omega )=\frac{1}{\sqrt{2\pi}}e^{-\Omega ^{2}/2},\;\;\;
\rho_{B}(\phi)=Z^{-1}e^{-U(\phi)}
\end{equation}
and $Z$ is  the partition function. 

To summarise, the differential equation (\ref{FP}) is the main result of section IID. The solution of equation (\ref{FP}) is used to compute $R(t)$ via (\ref{r2}), then  the  electronic transmission (\ref{trans}) is computed with help of $R(t)$ with subsequent calculations of electric current (\ref{JL-av3}).

\section{Results}

In this section,  numerical and analytical results are presented to illustrate the proposed theory. 
The rotational Langevin equations  (\ref{s1}) and (\ref{s2}) were conveniently written in terms of dimensionless units in which time is  measured in terms of  $\sqrt{I/(k_B T)}$. This means that rotational friction $\xi$ proportional to the solvent viscosity, which is an important parameter for the discussion  in the present section, is measured in units of thermal rotational frequency $\sqrt{k_B T/I}$. Whilst defining the Green's functions and solving Keldysh-Kadanoff-Baym equations we set $\hbar=1$, which means that level broadening $\Gamma$,  molecular orbital energy $\epsilon$, and electron-rotational coupling $\lambda$  will be also given in units of  $\sqrt{k_B T/I}$.
The Fermi energy $E_F$ of the leads is set to zero, hence  molecular orbital energy $\epsilon$ should be understood as the energy with respect to $E_F$. Since $\epsilon$ merely determines position of the transmission maximum, we set  
 $\epsilon$ throughout this section.  Without restricting the generality of the model this choice makes 
 discussions and analytical expressions more lucid and transparent for physical interpretation.
 
\begin{figure}
\centering
  \includegraphics[width=0.8 \linewidth]{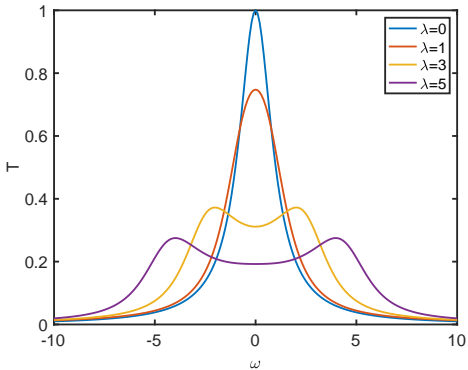}
  \caption{Electronic transmissions  $T(\omega)$ computed for various values of electron-rotational coupling $\lambda$. Parameters used in calculations: $\epsilon=0$, $\Gamma_L=\Gamma_R=\xi=1$.  The case of free rotation ($\alpha=0$) is considered here. }
  \label{fig:sg1}
\end{figure}

We begin with the consideration of free rotation, where the influence of the external alignment potential 
\begin{equation}\label{Uth}
U(\phi) = \alpha \cos (\phi - \phi_0)
\end{equation}
can be neglected ($\alpha=0$). 
Fig. \ref{fig:sg1} shows how the transmission scales with electron-rotational coupling $\lambda$.  The transmission is computed numerically
using algorithm detailed in Appendix A. 
For $\lambda=0$, electronic energy of the 
molecule is totally decoupled from its rotational motion. Hence the transmission is simply a Lorentzian centered at $\omega =  \epsilon$ with a width $\sim \Gamma$, as described by Eq. (\ref{Lor}). If  $\lambda$ increases, then Eq. (\ref{SLor}) predicts that the resonant value of the electronic transmission  decreases, $T(\omega=\epsilon) \sim \lambda^{-3/2}$, while the transmission starts to exhibit two symmetric maxima at $\omega = \epsilon\pm \omega_m$. Positions of these maxima obey the inequality $\omega_m = \sqrt{\lambda^2-\Gamma^2} < \lambda$ (see Eq. (\ref{Wm})) while the intensity of the transmission in the maxima scales as $T(\omega=\epsilon \pm \omega_m) \sim \lambda^{-1/2}$.   
 This behavior of the transmission can be rationalized by the consideration of the problem in the time domain:  
In the underdamped limit, according to Eq. (\ref{Jt}), $R(t) \sim J_0(\lambda t)$ and one can anticipate that the oscillation frequency  
$ \approx \lambda$ should manifest itself in the transmission for $ \lambda \gg \Gamma_{L}, \Gamma_{R}$. 

In terms of physical  observation Fig. \ref{fig:sg1} tells us that in the resonant regime, when molecular orbital energy is aligned with leads Fermi energy, the coupling to the surrounding solvent always suppresses the conductivity of the junction. However, quite oppositely,  if the electron transport is dominated by the molecular orbital which is shifted below or above Fermi energy,  the solvent  may considerably increase the conductance of the system.

\begin{figure}
	\centering
	\includegraphics[width=0.8 \linewidth]{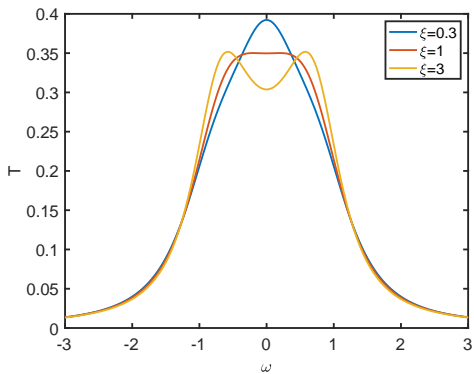}
	\caption{Electronic transmissions  $T(\omega)$  computed for various values of rotational friction $\xi$ in the case of  free rotation ($\alpha=0$). Parameters used in calculations: $\epsilon=0$, $\Gamma_L=\Gamma_R=0.3$, $\lambda=1$.  }
	\label{fig:sg3}
\end{figure}

There are three characteristic timescales in the system. One is associated with the time of electron tunneling across the molecule, $\Gamma^{-1}$. The other is the relaxation timescale for solvent dynamics, $\xi^{-1}$ in the underdamped limit and $\xi = D^{-1}$ in the overdamped limit ($D$ being the diffusion coefficient). The third, $\lambda^{-1}$, is associated with the electron-rotational coupling and, consequently, with the rate of energy exchange between tunneling electrons and external rotator.
Fig. \ref{fig:sg3} illustrates how  $T(\omega)$ is affected by the these timescales.  
As $\xi$ increases relative to $\Gamma$, $T(\omega)$ exhibits a transformation from a  single-peak to a  double-peak structure.
This somewhat counterintuitive behavior can be understood from the following considerations.  
In the limit $\lambda \gg 1$ and $\xi < \lambda$, the transmission is quite accurately described by the bath-free formula
of Eq. (\ref{Tw}), which has been analyzed  above. 
For $\lambda < 1$ and $\xi < 1$ (underdamped limit) $T(\omega)$ is also rather insensitive to $\xi$:  the rotational friction quenches slightly  the amplitude of oscillations in $R(t) \sim J_0(\lambda t)$, while the shape
 of $T(\omega)$ is largely determined by the coupling to the leads, through the exponentially decaying factor $\exp (-\Gamma \tau)$ as given in Eq. (\ref{trans}).
The situation changes in the overdamped limit ($\xi \gg \lambda$). In this case, $T(\omega)$ can be  evaluated analytically  (see Appendix \ref{Over}).
Eq. (\ref{con}) reveals then: the higher the friction, the more pronounced the peaks of $T(\omega)$. For $\lambda \ll 1$, for  example, Eq. (\ref{con}) reduces to 
\begin{equation}
\tilde{\rho}_{0}(s) \approx \frac{s+\xi^{-1}}{s^2+s\xi^{-1}+\lambda^2/2}.
\end{equation} 
The expression on the right has two simple poles at $(-\xi^{-1} \pm \sqrt{\xi^{-2}-2\lambda^2})/2$. For $\lambda > /(\sqrt{2}\xi)$ the poles become complex. Their imaginary part, which determines the $T(\omega)$ maxima $ \omega_m=\sqrt{2\lambda^2-\xi^{-2}}/2$, increases with friction, reaching a value of  
  $ \omega_m=\lambda/\sqrt{2} < \lambda$  in the overdamped limit $\xi \rightarrow \infty$. Then, if the broadening of the transmission induced by the coupling to the leads is not too high ($\Gamma < \omega_m$) or, equivalently, the  electron tunneling time  is not too short,  $T(\omega)$ exhibits the double-peak structure. In the context of the possibility to manipulate molecular conductivity, Fig. \ref{fig:sg3} reveals that the viscosity of the solvent can be used as effective control parameter for electron transport properties of molecular junction.

\begin{figure}
\centering
  \includegraphics[width=0.8 \linewidth]{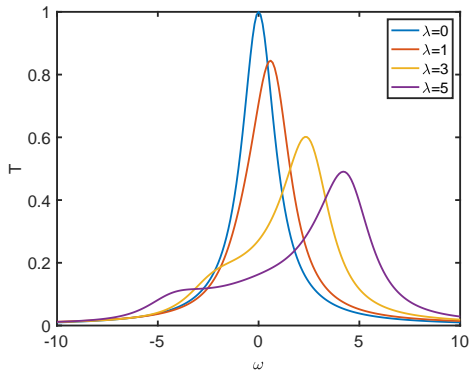}
  \caption{Electronic transmissions $T(\omega)$  computed for various values of electron-rotational coupling $\lambda$ in the presence of the external potential  $U(\phi)$  with $\alpha=1$ and $\phi_0 =0$. The remaining parameters:  
  	$\epsilon=0$, $\Gamma_L=\Gamma_R=1$, $\xi=1$.   }
  \label{fig:sg2}
\end{figure}

The effect of the external potential is illustrated by Fig. \ref{fig:sg2}, which shows  $T(\omega)$ for different $\lambda$ in the presence of the aligning potential (\ref{Uth}) with $\alpha=1$ and $\phi_0 =0$. Evidently, $U(\phi)$ breaks the rotational symmetry of the transmission, making the angles of $\pm \phi$ inequivalent. Hence the transmission develops particle-hole asymmetry, and  $T(\omega)$ has very different behavior  for hole  ($\omega<E_F$) and electron ($\omega>E_F$) transport.
The  particle-hole asymmetry becomes more profound as the strength of electron-rotation coupling $\lambda$ increases, producing a pronounced   maximum at positive $\omega$. As in the case of free rotation shown in Fig. \ref{fig:sg1}, the shift of the peak relative to the molecular orbital energy    is determined by  the electron-rotational coupling $\lambda$.  The asymmetry depends on the aligning angle $\phi_0$:  If $\phi = \pi$ is chosen, the peak of $T(\omega)$ is shifted to hole rather than electron region.

\section{conclusions}

In this paper, we have developed a quantum transport theory  which includes dynamical effects of solvent-molecule interaction in the non-equilibrium Green's function calculations of electric current.
The solvent is considered as  a classical macroscopic dipole moment which reorients stochastically and interacts with the electrons tunneling  through the molecular junctions.  This dynamical electron-rotation interaction makes the quantum mechanical Hamiltonian explicitly time-dependent. The  Keldysh-Kadanoff-Baym equations  are solved in time-domain. The obtained expression  for electric current requires the averaging over realisations of stochastic process related to the dipole moment rotation.  The  averaging over  random realisations of rotational variables is performed using Furutsu-Novikov method  to close hierarchy of equations for  stochastic correlation functions. 

We applied the theory for a model molecular junction - a single molecular orbital coupled to classical rotating dipole via electrostatic interaction.
Our calculations shows that, in the resonant regime when molecular orbital energy is aligned with the leads Fermi energy,  coupling to the surrounding solvent always suppresses the conductivity of the junction. However,   if the electron transport is controlled by the molecular orbital which is shifted below or above Fermi energy,  the solvent may noticeably  increase the conductance of the system. It is found that the viscosity of the solvent can be used as very effective stimulus to  control  electron transport properties of molecular junction.  If the rotation of the dipole moment is subjected to external aligning potential,  the solvent  breaks particle-hole symmetry of electron transport favouring holes or electron transport channel depending upon the aligning angle.

{{} 
{\bf Acknowledgment}
M.F.G. acknowledges support of Hangzhou Dianzi University through the startup funding.
}

\begin{center}
{\bf DATA AVAILABILITY}
\end{center}

The data that supports the findings of this study are available within the article.

\clearpage
\appendix

\section{Numerical method}

For obtaining a matrix form of (\ref{FP}), we use Hermit polynomials $H_n(\Omega)$ to expand
$\rho(\phi,\Omega ,t)$ in a series as 
\begin{equation}
\rho(\phi,\Omega ,t)=\sum_{k=-\infty}^{\infty}\,\sum_{n=0}^{\infty}e^{-ik\phi}g_{n}(\Omega )\rho_{kn}(t).
\end{equation}
where 
\begin{equation}
g_{n}(\Omega )=\frac{1}{n!}H_{n}(\Omega )\rho_{B}(\Omega ),
\end{equation}
and
\begin{equation}
\int_{-\infty}^{\infty}d\Omega  H_{m}(\Omega )g_{n}(\Omega )=\delta_{mn}.
\end{equation}
Then:
\begin{equation}
\int_{-\infty}^{\infty}d\Omega H_{m}(\Omega )\partial_{\Omega }g_{n}(\Omega )=-m\delta_{m-1,n}.
\end{equation}
\begin{equation}
\int_{-\infty}^{\infty}d\Omega H_{m}(\Omega )\Omega g_{n}(\Omega )=m\delta_{m-1,n}+\delta_{m+1,n}.
\end{equation}
\begin{equation}
\int_{-\infty}^{\infty}d\Omega H_{m}(\Omega )\left(\partial_{\Omega }\Omega +\partial_{\Omega }^{2}\right)g_{n}(\Omega )=-m\delta_{m,n}.
\end{equation}
Hence:
\begin{multline}
\partial_{t}\rho_{kn}(t)=-i\epsilon \rho_{kn}(t)+i\frac{\lambda}{2}\left[\rho_{k+1,n}(t)+\rho_{k-1,n}(t)\right] \\
+ik\left[\rho_{k,n+1}(t)+n\rho_{k,n-1}(t)\right]
\\
+i\frac{\alpha}{2}n\left[e^{-i\phi_0}\rho_{k+1,n-1}(t)-e^{i\phi_0}\rho_{k-1,n-1}(t)\right]-\xi n\rho_{kn}(t).\label{M}
\end{multline}
It should be solved with the initial condition 
\begin{equation}
\rho_{kn}(0)=\rho_{k}^{(0)}\delta_{n0}
\end{equation}
where 
\begin{equation}
\rho_{k}^{(0)}=\frac{1}{2\pi}\int_{0}^{2\pi}d\phi e^{ik\phi}\rho_{B}(\phi)
\end{equation}
Evidently,
 \begin{equation}
 R(t) =\rho_{00}(t).
 \end{equation}

 \section{Overdamped limit: Analytical expression for electronic transmission}\label{Over}

In the overdamped limit ($\xi \gg 1$) the Fokker-Planck equation (\ref{FP}) reduces to the rotational diffusion equation  
\begin{equation}
\partial_{t}\rho(\phi,t)=\left(-ih(\phi)+\xi^{-1}\left(\partial_{\phi}^{2}-\partial_{\phi}G(\phi)\right)\right)\rho(\phi,t).\label{FP1}
\end{equation}
In the case of free rotation ($G(\phi)=0$, $\alpha=0$) this  equation can be solved analytically.  
Expanding $\rho(\phi,t)$ in Fourier series, 
\begin{equation}
\rho(\phi,t)=\sum_{k=-\infty}^{\infty}e^{-ik\phi}\rho_{k}(t),
\end{equation}
we obtain 
\begin{equation}
\partial_{t}\rho_{k}(t)=-i\epsilon \rho_{k}(t)+i\frac{\lambda}{2}\left[\rho_{k+1}(t)+\rho_{k-1}(t)\right]-\xi^{-1}k^{2}\rho_{k}(t),
\label{r}
\end{equation}
\begin{equation}
\rho_{k}(0)=\delta_{0k}.
\end{equation}
We wish to evaluate 
\begin{equation}
R(t) =\rho_{0}(t).
\end{equation}
By introducing the Laplace transform 
\begin{equation}
\tilde{\rho}_{k}(s)=\int_0^\infty dt e^{-st}\rho_{k}(t), 
\end{equation}
we can rewrite  Eq. (\ref{r}) through the three-term recurrence relations 
\begin{equation}
-\delta_{0k}=a_{k}\tilde{\rho}_{k}(s)+b_{k}\tilde{\rho}_{k+1}(s)+c_{k}\tilde{\rho}_{k-1}(s)
\end{equation}
where 
\begin{equation}
a_{k}=-(s_\epsilon+\xi^{-1}k^{2}),\,\,b_{k}=c_{k}=i\frac{\lambda}{2}, \,\, s_\epsilon \equiv s + i  \epsilon.
\end{equation}
These recurrence relations can be solved through the continued fraction:
\begin{equation}\label{con}
\tilde{\rho}_{0}(s)=\cfrac{1}{s_\epsilon+\cfrac{\lambda^2/2}{s_\epsilon+\xi^{-1}+
		\cfrac{\lambda^2/4}{ s_\epsilon+4\xi^{-1} +\cfrac{\lambda^2/4}{s_\epsilon+9\xi^{-1}+...}}}}.
\end{equation}
Then, 
\begin{equation}
T(\omega)=\frac{1}{\pi}\frac{\Gamma_{L}\Gamma_{R}}{\Gamma_{L}+\Gamma_{R}}\mathrm{Re}\tilde{\rho}_{0}\left(i(\omega+\epsilon) +\Gamma\right).
\end{equation}

\section{Frictionless rigid rotor: analytical results for electronic transmission}\label{Free}

If $\xi=0$, Eq. (\ref{Rt}) can be evaluated analytically:
\begin{multline}\label{Jt}
R(t) =
e^{-i\epsilon t}\int_{-\infty}^{\infty}d\Omega \rho_{B}(\Omega ) \int_{0}^{2\pi} \frac{d \phi}{2 \pi} e^{i\lambda\int_{0}^{t}dt_1\cos(\phi+\Omega t_1)}
\\
=
e^{-i\epsilon t}\int_{0}^{2\pi} \frac{d \phi}{2 \pi} e^{i\lambda t\cos(\phi)}=
e^{-i\epsilon t}J_{0}(\lambda t),
\end{multline}
where $J_0(x)$ is the Bessel function of the first kind.
Since 
\begin{equation}
J_{0}(\lambda t)\approx\sqrt{\frac{2}{\pi\lambda t}}\cos(\lambda t-\pi/4)
\end{equation}
for large $t$, $R(t) $ exhibits algebraically damped oscillations with a period $2\pi/\lambda$
(for $\epsilon=0$). 

The transmission  of Eq. (\ref{trans})  can be evaluated as follows: 
\begin{multline}
T(\omega )=\mathrm{Re}\int_{0}^{\infty}dte^{-\Gamma t}e^{-i(\epsilon-\omega )t} \int_{0}^{2\pi}d\phi e^{i\lambda t\cos(\phi)}\\
=
\mathrm{Re}\int_{0}^{2\pi}d\phi\frac{1}{\Gamma+i(\epsilon-\omega -\lambda\cos(\phi))}
\\
=\frac{2 \pi }{\Gamma} \mathrm{Re} \frac{1}{\sqrt{\left[\Gamma+i(\epsilon-\omega )\right]^{2}+\lambda^{2}}}.
\end{multline}
	Explicitly,
\begin{equation}\label{Tw}
T(\omega )
=\frac{2 \pi }{\Gamma}  \sqrt{\frac{2\Gamma^2 - z + \sqrt{z^2 + 4 \lambda^2 \Gamma^2}}{2(z^2 + 4 \lambda^2 \Gamma^2)}}.
\end{equation}
where 
\begin{equation}\label{z}
z = (\epsilon-\omega)^2 + \Gamma^2 -\lambda^2 .
\end{equation}
In two particular cases, much simpler and easier-to-grasp formulas are deducible from Eq. (\ref{Tw}).  
If  $ \Gamma \gg \lambda$, the  transmission is described by the Lorentzian centered at $\omega =\epsilon$,
\begin{equation}\label{Lor}
T(\omega ) \sim  \frac{\Gamma}{(\epsilon-\omega)^2 + \Gamma^2}.
\end{equation}
In the opoosite limit $\lambda \gg \Gamma$, the transmission exhibits two symmetric maxima positioned at
$\epsilon -\omega = \pm \omega_m$:
\begin{equation}\label{SLor}
T(\omega ) \sim \sqrt{\frac{\lambda \Gamma}{((\epsilon-\omega)^2-\omega_m^2)^2 + 4 \lambda^2 \Gamma^2}}
\end{equation}
with
\begin{equation}\label{Wm}
\omega_m= \sqrt{\lambda^2 -\Gamma^2}.
\end{equation}

\clearpage

\begin{thebibliography}{20}
\expandafter\ifx\csname natexlab\endcsname\relax\def\natexlab#1{#1}\fi
\expandafter\ifx\csname bibnamefont\endcsname\relax
  \def\bibnamefont#1{#1}\fi
\expandafter\ifx\csname bibfnamefont\endcsname\relax
  \def\bibfnamefont#1{#1}\fi
\expandafter\ifx\csname citenamefont\endcsname\relax
  \def\citenamefont#1{#1}\fi
\expandafter\ifx\csname url\endcsname\relax
  \def\url#1{\texttt{#1}}\fi
\expandafter\ifx\csname urlprefix\endcsname\relax\def\urlprefix{URL }\fi
\providecommand{\bibinfo}[2]{#2}
\providecommand{\eprint}[2][]{\url{#2}}

\bibitem[{\citenamefont{Xiang et~al.}(2016)\citenamefont{Xiang, Wang, Jia, Lee,
  and Guo}}]{doi:10.1021/acs.chemrev.5b00680}
\bibinfo{author}{\bibfnamefont{D.}~\bibnamefont{Xiang}},
  \bibinfo{author}{\bibfnamefont{X.}~\bibnamefont{Wang}},
  \bibinfo{author}{\bibfnamefont{C.}~\bibnamefont{Jia}},
  \bibinfo{author}{\bibfnamefont{T.}~\bibnamefont{Lee}}, \bibnamefont{and}
  \bibinfo{author}{\bibfnamefont{X.}~\bibnamefont{Guo}},
  \bibinfo{journal}{Chemical Reviews} \textbf{\bibinfo{volume}{116}},
  \bibinfo{pages}{4318} (\bibinfo{year}{2016}), \bibinfo{note}{pMID: 26979510},
  \eprint{https://doi.org/10.1021/acs.chemrev.5b00680},
  \urlprefix\url{https://doi.org/10.1021/acs.chemrev.5b00680}.

\bibitem[{\citenamefont{Wu et~al.}(2008)\citenamefont{Wu, Gonz{\'a}lez, Huber,
  Grunder, Mayor, Sch{\"o}nenberger, and Calame}}]{Wu:2008aa}
\bibinfo{author}{\bibfnamefont{S.}~\bibnamefont{Wu}},
  \bibinfo{author}{\bibfnamefont{M.~T.} \bibnamefont{Gonz{\'a}lez}},
  \bibinfo{author}{\bibfnamefont{R.}~\bibnamefont{Huber}},
  \bibinfo{author}{\bibfnamefont{S.}~\bibnamefont{Grunder}},
  \bibinfo{author}{\bibfnamefont{M.}~\bibnamefont{Mayor}},
  \bibinfo{author}{\bibfnamefont{C.}~\bibnamefont{Sch{\"o}nenberger}},
  \bibnamefont{and} \bibinfo{author}{\bibfnamefont{M.}~\bibnamefont{Calame}},
  \bibinfo{journal}{Nature Nanotechnology} \textbf{\bibinfo{volume}{3}},
  \bibinfo{pages}{569} (\bibinfo{year}{2008}),
  \urlprefix\url{https://doi.org/10.1038/nnano.2008.237}.

\bibitem[{\citenamefont{Nishino et~al.}(2013)\citenamefont{Nishino, Hayashi,
  and Bui}}]{doi:10.1021/ja311463b}
\bibinfo{author}{\bibfnamefont{T.}~\bibnamefont{Nishino}},
  \bibinfo{author}{\bibfnamefont{N.}~\bibnamefont{Hayashi}}, \bibnamefont{and}
  \bibinfo{author}{\bibfnamefont{P.~T.} \bibnamefont{Bui}},
  \bibinfo{journal}{Journal of the American Chemical Society}
  \textbf{\bibinfo{volume}{135}}, \bibinfo{pages}{4592} (\bibinfo{year}{2013}),
  \bibinfo{note}{pMID: 23488642}, \eprint{https://doi.org/10.1021/ja311463b},
  \urlprefix\url{https://doi.org/10.1021/ja311463b}.

\bibitem[{\citenamefont{Bui et~al.}(2013)\citenamefont{Bui, Nishino, Yamamoto,
  and Shiigi}}]{doi:10.1021/ja312019p}
\bibinfo{author}{\bibfnamefont{P.~T.} \bibnamefont{Bui}},
  \bibinfo{author}{\bibfnamefont{T.}~\bibnamefont{Nishino}},
  \bibinfo{author}{\bibfnamefont{Y.}~\bibnamefont{Yamamoto}}, \bibnamefont{and}
  \bibinfo{author}{\bibfnamefont{H.}~\bibnamefont{Shiigi}},
  \bibinfo{journal}{Journal of the American Chemical Society}
  \textbf{\bibinfo{volume}{135}}, \bibinfo{pages}{5238} (\bibinfo{year}{2013}),
  \bibinfo{note}{pMID: 23534478}, \eprint{https://doi.org/10.1021/ja312019p},
  \urlprefix\url{https://doi.org/10.1021/ja312019p}.

\bibitem[{\citenamefont{Kuznetsov}(2007)}]{kuznetsov-ndr}
\bibinfo{author}{\bibfnamefont{A.~M.} \bibnamefont{Kuznetsov}},
  \bibinfo{journal}{J. Chem. Phys.} \textbf{\bibinfo{volume}{127}},
  \bibinfo{pages}{084710} (\bibinfo{year}{2007}).

\bibitem[{\citenamefont{Milan et~al.}(2016)\citenamefont{Milan, Al-Owaedi,
  Oerthel, Marqu{\'e}s-Gonz{\'a}lez, Brooke, Bryce, Cea, Ferrer, Higgins,
  Lambert et~al.}}]{doi:10.1021/acs.jpcc.5b08877}
\bibinfo{author}{\bibfnamefont{D.~C.} \bibnamefont{Milan}},
  \bibinfo{author}{\bibfnamefont{O.~A.} \bibnamefont{Al-Owaedi}},
  \bibinfo{author}{\bibfnamefont{M.-C.} \bibnamefont{Oerthel}},
  \bibinfo{author}{\bibfnamefont{S.}~\bibnamefont{Marqu{\'e}s-Gonz{\'a}lez}},
  \bibinfo{author}{\bibfnamefont{R.~J.} \bibnamefont{Brooke}},
  \bibinfo{author}{\bibfnamefont{M.~R.} \bibnamefont{Bryce}},
  \bibinfo{author}{\bibfnamefont{P.}~\bibnamefont{Cea}},
  \bibinfo{author}{\bibfnamefont{J.}~\bibnamefont{Ferrer}},
  \bibinfo{author}{\bibfnamefont{S.~J.} \bibnamefont{Higgins}},
  \bibinfo{author}{\bibfnamefont{C.~J.} \bibnamefont{Lambert}},
  \bibnamefont{et~al.}, \bibinfo{journal}{The Journal of Physical Chemistry C}
  \textbf{\bibinfo{volume}{120}}, \bibinfo{pages}{15666}
  (\bibinfo{year}{2016}), \eprint{https://doi.org/10.1021/acs.jpcc.5b08877},
  \urlprefix\url{https://doi.org/10.1021/acs.jpcc.5b08877}.

\bibitem[{\citenamefont{Dzhioev and Kosov}(2012)}]{solvent12}
\bibinfo{author}{\bibfnamefont{A.~A.} \bibnamefont{Dzhioev}} \bibnamefont{and}
  \bibinfo{author}{\bibfnamefont{D.~S.} \bibnamefont{Kosov}},
  \bibinfo{journal}{Phys. Rev. B} \textbf{\bibinfo{volume}{85}},
  \bibinfo{pages}{033408} (\bibinfo{year}{2012}),
  \urlprefix\url{https://link.aps.org/doi/10.1103/PhysRevB.85.033408}.

\bibitem[{\citenamefont{Leary et~al.}(2009)\citenamefont{Leary, H\"obenreich,
  Higgins, van Zalinge, Haiss, Nichols, Finch, Grace, Lambert, McGrath
  et~al.}}]{PhysRevLett.102.086801}
\bibinfo{author}{\bibfnamefont{E.}~\bibnamefont{Leary}},
  \bibinfo{author}{\bibfnamefont{H.}~\bibnamefont{H\"obenreich}},
  \bibinfo{author}{\bibfnamefont{S.~J.} \bibnamefont{Higgins}},
  \bibinfo{author}{\bibfnamefont{H.}~\bibnamefont{van Zalinge}},
  \bibinfo{author}{\bibfnamefont{W.}~\bibnamefont{Haiss}},
  \bibinfo{author}{\bibfnamefont{R.~J.} \bibnamefont{Nichols}},
  \bibinfo{author}{\bibfnamefont{C.~M.} \bibnamefont{Finch}},
  \bibinfo{author}{\bibfnamefont{I.}~\bibnamefont{Grace}},
  \bibinfo{author}{\bibfnamefont{C.~J.} \bibnamefont{Lambert}},
  \bibinfo{author}{\bibfnamefont{R.}~\bibnamefont{McGrath}},
  \bibnamefont{et~al.}, \bibinfo{journal}{Phys. Rev. Lett.}
  \textbf{\bibinfo{volume}{102}}, \bibinfo{pages}{086801}
  (\bibinfo{year}{2009}),
  \urlprefix\url{https://link.aps.org/doi/10.1103/PhysRevLett.102.086801}.

\bibitem[{\citenamefont{Stipe et~al.}(1997)\citenamefont{Stipe, Rezaei, Ho,
  Gao, Persson, and Lundqvist}}]{PhysRevLett.78.4410}
\bibinfo{author}{\bibfnamefont{B.~C.} \bibnamefont{Stipe}},
  \bibinfo{author}{\bibfnamefont{M.~A.} \bibnamefont{Rezaei}},
  \bibinfo{author}{\bibfnamefont{W.}~\bibnamefont{Ho}},
  \bibinfo{author}{\bibfnamefont{S.}~\bibnamefont{Gao}},
  \bibinfo{author}{\bibfnamefont{M.}~\bibnamefont{Persson}}, \bibnamefont{and}
  \bibinfo{author}{\bibfnamefont{B.~I.} \bibnamefont{Lundqvist}},
  \bibinfo{journal}{Phys. Rev. Lett.} \textbf{\bibinfo{volume}{78}},
  \bibinfo{pages}{4410} (\bibinfo{year}{1997}).

\bibitem[{\citenamefont{Luka-Guth et~al.}(2016)\citenamefont{Luka-Guth,
  Hambsch, Bloch, Ehrenreich, Briechle, Kilibarda, Sendler, Sysoiev, Huhn, Erbe
  et~al.}}]{solvent2016}
\bibinfo{author}{\bibfnamefont{K.}~\bibnamefont{Luka-Guth}},
  \bibinfo{author}{\bibfnamefont{S.}~\bibnamefont{Hambsch}},
  \bibinfo{author}{\bibfnamefont{A.}~\bibnamefont{Bloch}},
  \bibinfo{author}{\bibfnamefont{P.}~\bibnamefont{Ehrenreich}},
  \bibinfo{author}{\bibfnamefont{B.~M.} \bibnamefont{Briechle}},
  \bibinfo{author}{\bibfnamefont{F.}~\bibnamefont{Kilibarda}},
  \bibinfo{author}{\bibfnamefont{T.}~\bibnamefont{Sendler}},
  \bibinfo{author}{\bibfnamefont{D.}~\bibnamefont{Sysoiev}},
  \bibinfo{author}{\bibfnamefont{T.}~\bibnamefont{Huhn}},
  \bibinfo{author}{\bibfnamefont{A.}~\bibnamefont{Erbe}}, \bibnamefont{et~al.},
  \bibinfo{journal}{Beilstein J. Nanotechnol.} \textbf{\bibinfo{volume}{7}},
  \bibinfo{pages}{1055} (\bibinfo{year}{2016}).

\bibitem[{\citenamefont{Kotiuga et~al.}(2015)\citenamefont{Kotiuga, Darancet,
  Arroyo, Venkataraman, and Neaton}}]{Kotiuga:2015aa}
\bibinfo{author}{\bibfnamefont{M.}~\bibnamefont{Kotiuga}},
  \bibinfo{author}{\bibfnamefont{P.}~\bibnamefont{Darancet}},
  \bibinfo{author}{\bibfnamefont{C.~R.} \bibnamefont{Arroyo}},
  \bibinfo{author}{\bibfnamefont{L.}~\bibnamefont{Venkataraman}},
  \bibnamefont{and} \bibinfo{author}{\bibfnamefont{J.~B.}
  \bibnamefont{Neaton}}, \bibinfo{journal}{Nano Letters}
  \textbf{\bibinfo{volume}{15}}, \bibinfo{pages}{4498} (\bibinfo{year}{2015}),
  \urlprefix\url{https://doi.org/10.1021/acs.nanolett.5b00990}.

\bibitem[{\citenamefont{Fatemi et~al.}(2011)\citenamefont{Fatemi, Kamenetska,
  Neaton, and Venkataraman}}]{doi:10.1021/nl200324e}
\bibinfo{author}{\bibfnamefont{V.}~\bibnamefont{Fatemi}},
  \bibinfo{author}{\bibfnamefont{M.}~\bibnamefont{Kamenetska}},
  \bibinfo{author}{\bibfnamefont{J.~B.} \bibnamefont{Neaton}},
  \bibnamefont{and}
  \bibinfo{author}{\bibfnamefont{L.}~\bibnamefont{Venkataraman}},
  \bibinfo{journal}{Nano Letters} \textbf{\bibinfo{volume}{11}},
  \bibinfo{pages}{1988} (\bibinfo{year}{2011}), \bibinfo{note}{pMID: 21500833},
  \eprint{https://doi.org/10.1021/nl200324e},
  \urlprefix\url{https://doi.org/10.1021/nl200324e}.

\bibitem[{\citenamefont{Ghane et~al.}(2015)\citenamefont{Ghane, Kleshchonok,
  Gutierrez, and Cuniberti}}]{doi:10.1021/acs.jpcc.5b06867}
\bibinfo{author}{\bibfnamefont{T.}~\bibnamefont{Ghane}},
  \bibinfo{author}{\bibfnamefont{A.}~\bibnamefont{Kleshchonok}},
  \bibinfo{author}{\bibfnamefont{R.}~\bibnamefont{Gutierrez}},
  \bibnamefont{and}
  \bibinfo{author}{\bibfnamefont{G.}~\bibnamefont{Cuniberti}},
  \bibinfo{journal}{The Journal of Physical Chemistry C}
  \textbf{\bibinfo{volume}{119}}, \bibinfo{pages}{20201}
  (\bibinfo{year}{2015}), \eprint{https://doi.org/10.1021/acs.jpcc.5b06867},
  \urlprefix\url{https://doi.org/10.1021/acs.jpcc.5b06867}.

\bibitem[{\citenamefont{Choi et~al.}(2016)\citenamefont{Choi, Capozzi, Ahn,
  Turkiewicz, Lovat, Nuckolls, Steigerwald, Venkataraman, and
  Roy}}]{C5SC02595H}
\bibinfo{author}{\bibfnamefont{B.}~\bibnamefont{Choi}},
  \bibinfo{author}{\bibfnamefont{B.}~\bibnamefont{Capozzi}},
  \bibinfo{author}{\bibfnamefont{S.}~\bibnamefont{Ahn}},
  \bibinfo{author}{\bibfnamefont{A.}~\bibnamefont{Turkiewicz}},
  \bibinfo{author}{\bibfnamefont{G.}~\bibnamefont{Lovat}},
  \bibinfo{author}{\bibfnamefont{C.}~\bibnamefont{Nuckolls}},
  \bibinfo{author}{\bibfnamefont{M.~L.} \bibnamefont{Steigerwald}},
  \bibinfo{author}{\bibfnamefont{L.}~\bibnamefont{Venkataraman}},
  \bibnamefont{and} \bibinfo{author}{\bibfnamefont{X.}~\bibnamefont{Roy}},
  \bibinfo{journal}{Chem. Sci.} \textbf{\bibinfo{volume}{7}},
  \bibinfo{pages}{2701} (\bibinfo{year}{2016}),
  \urlprefix\url{http://dx.doi.org/10.1039/C5SC02595H}.

\bibitem[{\citenamefont{Kornyshev et~al.}(2006)\citenamefont{Kornyshev,
  Kuznetsov, and Ulstrup}}]{Kornyshev6799}
\bibinfo{author}{\bibfnamefont{A.~A.} \bibnamefont{Kornyshev}},
  \bibinfo{author}{\bibfnamefont{A.~M.} \bibnamefont{Kuznetsov}},
  \bibnamefont{and} \bibinfo{author}{\bibfnamefont{J.}~\bibnamefont{Ulstrup}},
  \bibinfo{journal}{Proceedings of the National Academy of Sciences}
  \textbf{\bibinfo{volume}{103}}, \bibinfo{pages}{6799} (\bibinfo{year}{2006}),
  \eprint{https://www.pnas.org/content/103/18/6799.full.pdf},
  \urlprefix\url{https://www.pnas.org/content/103/18/6799}.

\bibitem[{\citenamefont{Zhang et~al.}(2008)\citenamefont{Zhang, Kuznetsov,
  Medvedev, Chi, Albrecht, Jensen, and Ulstrup}}]{doi:10.1021/cr068073+}
\bibinfo{author}{\bibfnamefont{J.}~\bibnamefont{Zhang}},
  \bibinfo{author}{\bibfnamefont{A.~M.} \bibnamefont{Kuznetsov}},
  \bibinfo{author}{\bibfnamefont{I.~G.} \bibnamefont{Medvedev}},
  \bibinfo{author}{\bibfnamefont{Q.}~\bibnamefont{Chi}},
  \bibinfo{author}{\bibfnamefont{T.}~\bibnamefont{Albrecht}},
  \bibinfo{author}{\bibfnamefont{P.~S.} \bibnamefont{Jensen}},
  \bibnamefont{and} \bibinfo{author}{\bibfnamefont{J.}~\bibnamefont{Ulstrup}},
  \bibinfo{journal}{Chemical Reviews} \textbf{\bibinfo{volume}{108}},
  \bibinfo{pages}{2737} (\bibinfo{year}{2008}), \bibinfo{note}{pMID: 18620372},
  \eprint{https://doi.org/10.1021/cr068073+},
  \urlprefix\url{https://doi.org/10.1021/cr068073+}.

\bibitem[{\citenamefont{Klyatskin}(2005)}]{klyatskin2005}
\bibinfo{author}{\bibfnamefont{V.~I.} \bibnamefont{Klyatskin}},
  \emph{\bibinfo{title}{Dynamics of Stochastic Systems}}
  (\bibinfo{publisher}{Amsterdam: Elsevier}, \bibinfo{year}{2005}).

\bibitem[{\citenamefont{Coffey et~al.}(1996)\citenamefont{Coffey, Kalmykov, and
  Waldron}}]{coffey}
\bibinfo{author}{\bibfnamefont{W.~T.} \bibnamefont{Coffey}},
  \bibinfo{author}{\bibfnamefont{Y.~P.} \bibnamefont{Kalmykov}},
  \bibnamefont{and} \bibinfo{author}{\bibfnamefont{J.}~\bibnamefont{Waldron}},
  \emph{\bibinfo{title}{The Langevin Equation: With Applications in Physics,
  Chemistry and Electrical Engineering}} (\bibinfo{publisher}{World Scientific,
  Singapore}, \bibinfo{year}{1996}).

\bibitem[{\citenamefont{Haug and Jauho}(2010)}]{haug-jauho}
\bibinfo{author}{\bibfnamefont{H.}~\bibnamefont{Haug}} \bibnamefont{and}
  \bibinfo{author}{\bibfnamefont{A.}~\bibnamefont{Jauho}},
  \emph{\bibinfo{title}{Quantum Kinetics in Transport and Optics of
  Semiconductors}} (\bibinfo{publisher}{Springer, Berlin/Heidelberg},
  \bibinfo{year}{2010}).

\bibitem[{\citenamefont{Van~Kampen}(2007)}]{vanKampen}
\bibinfo{author}{\bibfnamefont{N.~G.} \bibnamefont{Van~Kampen}},
  \emph{\bibinfo{title}{Stochastic Processes in Physics and Chemistry}}
  (\bibinfo{publisher}{3th ed. Amsterdam: North-Holland Personal Library},
  \bibinfo{year}{2007}).

\end{thebibliography}

\end{document}